\documentclass{article}
\usepackage{graphicx} 

\title{Dynamic PES with MPEC}
\author{Alice Xuming Feng\thanks{Department of Management Science and Engineering, Stanford University.},
  Kenneth L. Judd\thanks{Hoover Institution, Stanford University.}}
\date{September 2026}
\usepackage[letterpaper,top=2cm,bottom=2cm,left=2.5cm,right=2.5cm,marginparwidth=1.75cm]{geometry}
\usepackage{amsmath}
\usepackage{amsfonts}
\usepackage{booktabs}
\usepackage{subcaption}
\usepackage{flafter} 
\usepackage[section]{placeins} 
\usepackage{float}
\usepackage[round,authoryear]{natbib}
\begin{document}
\maketitle

\begin{abstract}
We study the design of dynamic contracts for payments for ecosystem services (PES), allowing for flexible payment structures within an infinite-horizon principal–agent framework. We formulate the optimal contract design problem as a mathematical program with equilibrium constraints (MPEC), enabling the numerical characterization of optimal payment schedules. Our experiments examine how optimal contracts depend on the parties' discount factors, environmental depreciation, and the functional form of the principal's environmental benefits. Despite the flexibility of the contract space, optimal payment schedules across many specifications exhibit a simple structure: payments begin only after a certain level of environmental quality is achieved and increase approximately linearly thereafter. These findings offer insights into the design of simple and effective dynamic PES contracts.
\end{abstract}

\section{Introduction}
Payments for ecosystem services have become an important policy instrument for
forest conservation, watershed protection, carbon sequestration, and ecological
restoration. Their central idea is straightforward: environmental services
often generate benefits for people other than the landholder who controls the
relevant resource, so private incentives may lead to too little conservation
or restoration. A PES program attempts to internalize this externality by
making compensation conditional on a desired environmental outcome or on an
observable land-use practice that is expected to produce that outcome
\citep{wunder2005,engel2008}. Following the PES literature, we use \textit{provider} to refer to the landholder or resource manager whose actions generate the ecosystem service. In our formal principal–agent model, the provider is the agent.

Implementing this idea is more difficult than the basic logic suggests.
Environmental outcomes are heterogeneous across space, costly to monitor. Improvements on contracted land can
also displace harmful activity elsewhere, as documented in Mexico's national
PES program \citep{alixgarcia2012}. Moreover, many environmental investments
are durable assets whose quality can improve with sustained effort but can also
depreciate through fire, erosion, illegal clearing, or the abandonment of
maintenance. These features make PES a dynamic contracting problem rather than
simply a one-period subsidy problem.

This paper studies a dynamic PES contract when environmental quality is
observed by both parties but the agent's effort is hidden. In each period, the
agent chooses effort that shifts the distribution of future environmental
quality. The principal observes the current state $\theta$ and pays according to a
payment schedule $\tau(\theta)$. We restrict attention to
payments based on contemporaneous environmental quality rather than the full
history. This restriction reflects the administrative simplicity of many
government programs and allows us to examine how the shape of the payment
schedule affects effort throughout the state space.

The model focuses on three ingredients. First, it allows the principal and
agent to discount the future differently. This matters because future
state-contingent payments are less effective when the agent is impatient.
Second, it incorporates stochastic depreciation, so environmental progress is
not necessarily permanent. Third, it permits an arbitrary monotone payment
schedule and therefore allows a direct numerical comparison between a flexible
nonlinear contract and a restricted linear contract. After discretizing the
state space, the principal's problem is written as a mathematical program with
equilibrium constraints (MPEC), with the agent's Bellman equation and
first-order incentive condition imposed as constraints.

The numerical results reveal several economically intuitive and quantitatively
important patterns. When the agent becomes more patient, the principal will
promise lower current payments while inducing greater effort over most states.
Introducing depreciation increases the optimal payment to the agent  and increases agent's corresponding effort level, while leaving the shape of optimal contract largely unchanged.
Concave environmental benefits to the principal shift incentives toward lower and intermediate
states and reduce effort near the upper part of the state space relative to
linear benefits. Finally, in the baseline calibration, the saved numerical
solutions suggest that a nonlinear schedule targets incentives differently
from a restricted linear schedule.

The remainder of the paper proceeds as follows. Section 2 reviews the PES and
contract-design literature. Section 3 presents the dynamic model, including
the depreciation extension, and Section 4 reports the numerical experiments.
The appendix describes the capped Poisson implementation.

\section{Related literature}
\subsection{PES principles, design, and empirical performance}
The modern PES literature begins with the definition in \citet{wunder2005},
which emphasizes a voluntary transaction, a defined environmental service or
associated land use, a buyer and provider, and payment conditional on service
provision. \citet{engel2008} place this definition in an environmental
economics framework: PES can internalize external benefits when ecosystem
managers do not capture the full social value of conservation. They also
emphasize that PES is not universally appropriate; effectiveness depends on
the underlying market failure, property rights, monitoring feasibility, and
program design.

Comparative evidence confirms that design choices matter.
\citet{wundertakingstock2008} find that user-financed programs tend to be more
targeted, locally tailored, closely monitored, and willing to enforce
conditionality than government-financed programs, which often pursue several
side objectives. Reviewing a wider empirical literature,
\citet{borner2017} conclude that PES effectiveness often falls short of early
expectations and depends on targeting, payment differentiation, contract
length, transaction costs, additionality, and spillovers.

Two empirical concerns are especially relevant for contract design. First,
providers may opt into programs even when their behavior would have
been environmentally desirable without payment. Enrollment costs and
administrative burdens can change both participation and the composition of
participants \citep{jack2019}. Second, environmental gains can leak outside
the contracted area. \citet{alixgarcia2012} estimate that Mexico's program
reduced clearing on enrolled parcels, but low baseline deforestation limited
total additionality and slippage reduced effectiveness. These findings motivate
contracts that reward genuine state improvements and account for transitions
throughout the program.

\subsection{Permanence, poverty, and equity}
PES programs frequently combine environmental objectives with livelihood or
poverty goals. \citet{wunderpoor2008} finds that poor providers can often
participate and benefit, but argues that the aggregate poverty effects are
generally modest and that PES should retain a primary environmental focus.
Other work stresses that efficiency and equity cannot always be separated.
\citet{pascual2010} show how institutions, bargaining power, perceptions of
fairness, and the distribution of payments can affect participation and
environmental performance. \citet{muradian2013} therefore caution against
treating PES as an automatic win--win solution or as a purely technical market
instrument detached from political and social context.

Permanence is particularly important for restoration or ``asset-building''
programs, where payments may end after the environmental asset has improved.
Evidence from a silvopastoral program in Nicaragua shows that productive
land-use changes can persist after payments cease \citep{pagiola2020}. This
evidence suggests that the dynamics of the environmental asset and the
agent's private return from maintaining it are central to contract design. Our
model captures a related issue from the opposite direction: even while a
program continues, stochastic depreciation can undo part of the accumulated
environmental improvement and alter the value of incentives.

\subsection{Contract theory and PES}
A growing theoretical literature studies PES through principal--agent and
market-design models. \citet{mason2013} analyze adverse selection in carbon
offsets and show how menus of contracts can reduce information rents and
identify additional forest conservation. \citet{teytelboym2019} emphasizes
that natural-capital markets are complicated by heterogeneous assets, high
monitoring and transaction costs, incomplete property rights, and ecological
complementarities.

Most closely related to our contract comparison,
\citet{daili2023} study a static hidden-information problem and show that a
simple linear payment can guarantee at least half of the optimal contract
payoff without knowledge of the distribution of baseline conservation.
\citet{lomurthy2025} extend the analysis to an infinite-horizon asset-building
setting with noisy observations and hidden action. They study additionality,
permanence, and poverty alleviation and derive approximation guarantees for
linear contracts.

\section{Setting}
We analyze how a government policymaker can create incentives for people to
improve environmental quality. Our dynamic framework is related
to \citet{lomurthy2025}, but we assume a different information structure and
contract space.

We assume a single principal with many agents. Each agent has a type $\theta$. Both the principal and agent observe the agent's type. The agent controls his type, $\theta$, and his effort can affect the type. In each period, an agent chooses an effort m which may cause a change in the type. Formally,  $\theta_{t+1}-\theta_t\sim X$ where the pdf $f_X(x)=p(x;m).$ The support for X can include some negative values, representing depreciation. In each period, the principal pays the agent an amount $\tau$ which depends on $\theta$. 

This is not a full dynamic principal agent analysis. In many dynamic principal agent problems, the principal remembers the history of information about the agent and uses the full history as well as any current information to determine payments to the agent. Government policies, such as income taxation, are usually less complex with payments depending only on contemporaneous information. 
This is the same approach taken in Lo-Murthy. but we allow the payment $\tau$ to be an arbitrary function of $\theta$. This allows us to give a more refined analysis of what determines the optimal payment policy.

The principal and agent both aim to maximize the expected present value of their utility. They have possibly different discount rates denoted by $\beta_p$ and $\beta_a$. In general, the principal is more patient than the agent. 

Our model differs from the Lo-Murthy model, but shares important elements. $\theta$ is the state, is observed by both principal and agent, and determines the payment to the agent. We assume that the principal observes the state $\theta$ with no error whereas Lo-Murthy assumes some measurement error. The presence of measurement error greatly complicates the problem. We choose to first examine the perfect observation case to establish some results. 

If there is no depreciation, then there is an absorbing state $\bar{\theta}$ like the assumption in Lo-Murthy. The lumpsum payment in their paper is equivalent to the perpetuity $\tau(\bar{\theta})$ in our paper

We use $\tau(\theta)$ to denote the payment contract. The agents' bellman equation is 
$$V_a(\theta)=\max\limits_{m}\tau(\theta)-c(m)+\beta_a\mathbb{E}[V_a(\theta^+)|m]$$
In the meantime, the value function for the principal satisfies this recursive definition
$$V_p=u_p(\theta)-\tau(\theta)+\beta_p\mathbb{E}[V_p(\theta^+)|m]$$
In general, the principal seeks to maximum the expected value at time zero: if the distribution of agent's initial type has cdf $f(\theta)$, then the principal's objective function becomes $$\int_0^1V_p(\theta)f(\theta)d\theta$$
We consider two versions of the state dynamics. The benchmark case has no
depreciation, as in \citet{lomurthy2025}, so the upper state is absorbing. We
then add stochastic depreciation, allowing the environmental state to fall
after it has improved.
Numerically, we discretize $\theta$ into n+1 values, with $\theta_{(i)}=\frac{i\bar{\theta}}{n}$. Define $v^a_i=V_a(\theta_i)$ and $v^p_i=V_p(\theta_i)$.  We use $m_i$ to denote the agent's optimal effort at state $\theta_i$. Following an effort $m$, the transition probability from $\theta_{i}$ to $\theta_{i+j}$ follows some probability distribution $p_{ij}(m)$

We can then rewrite the optimal contract problem as a constraint nonlinear optimization with equality constraint problem.
$$\max\limits_{\{\tau_i\},\{v^p_i\},\{v^a_i\},\{m_i\}} v^p_0-\sum\limits_{i=0}^{n-1}\mu(\tau_{i+1}-\tau_{i})^2$$
s.t\\
For $i=0,..,n-1$
$$v^a_i=\tau_i-c(m_i)+\beta_a (\sum\limits_{j=0}^{n-i} v^a_{i+j}p_{ij}(m_i))$$
$$v^p_i=u_p(\frac{i\bar{\theta}}{n})-\tau_i+\beta_p (\sum\limits_{j=0}^{n-i} v^p_{i+j}p_{ij}(m_i))$$ 
$$0=-c'(m_i)+\beta_a (\sum\limits_{j=0}^{n-i} v^a_{i+j}p_{ij}'(m_i))$$
$$-c''(m_i)+\beta_a (\sum\limits_{j=0}^{n-i} v^a_{i+j}p_{ij}''(m_i))\le 0$$
For $i=n$
$$v^a_n=\frac{\tau_n}{1-\beta_a}$$
$$v^p_n=\frac{u_p(\bar{\theta})-\tau_n}{1-\beta_p}$$
$$m_n=0$$
Additionally, we require
\[
\begin{aligned}
\tau_i &\geq 0, && i=0,\ldots,n,\\
\tau_{i+1} &\geq\tau_i, && i=0,\ldots,n-1.
\end{aligned}
\]
These constraints reflect a PES program that rewards better environmental
outcomes: payment cannot be negative, and higher environmental quality cannot
lead to a lower payment. To stabilize the numerical optimization, we also
include a payment-smoothing penalty with weight $\mu$. Ideally, we would set
$\mu=0$, but the resulting problem is highly ill-conditioned, so we use a
small positive value.

When solving the MPEC problem, the second-order conditions are not imposed. Instead, they are checked afterwards.
\subsection{Adding depreciation to the model}
Suppose that, after the upward transition generated by effort, the state
depreciates with probability $q$. When depreciation occurs, the state falls by
$\delta$ grid points, or $\delta\bar{\theta}/n$ in state units. The dynamic
programming MPEC becomes
$$\max\limits_{\{\tau_i\},\{v^p_i\},\{v^a_i\},\{m_i\}} v^p_0-\sum\limits_{i=0}^{n-1}\mu(\tau_{i+1}-\tau_{i})^2$$
s.t\\
For $i=0,..,n$
$$v^a_i=\tau_i-c(m_i)+\beta_a \sum\limits_{j=0}^{n-i} ((1-q)v^a_{i+j}p_{ij}(m_i)+q v^a_{[i+j-\delta]^+}p_{ij}(m_i))$$
$$v^p_i=u_p(\frac{i\bar{\theta}}{n})-\tau_i+\beta_p \sum\limits_{j=0}^{n-i}((1-q)v^p_{i+j}p_{ij}(m_i)+qv^p_{[i+j-\delta]^+}p_{ij}(m_i))$$ 
$$0=-c'(m)+\beta_a \sum\limits_{j=0}^{n-i} ((1-q)v^a_{i+j}p_{ij}'(m_i)+q v^a_{[i+j-\delta]^+}p_{ij}'(m_i))$$
\[
-c''(m_i)
+
\beta_a
\sum_{j=0}^{n-i}
\left(
(1-q)v^a_{i+j}
+
qv^a_{[i+j-\delta]_+}
\right)
p''_{ij}(m_i)
\le 0.
\]
\section{Numerical experiments}
We next study a few numerical examples. The purpose is to demonstrate the tractability of our approach and highlight the qualitative features of the optimal payment scheme. 

In the numerical experiments, the principal's baseline utility is linear,
$u_p(\theta)=k\theta$, and the agent's effort cost is quadratic,
$c(m)=hm^2$. Unless otherwise stated, all experiments use this linear principal
utility function. All agents start at $\theta_0=0$. The computations use a capped
Poisson transition to map effort into stochastic upward movement on the state
grid. The attempted upward movement is a Poisson random variable with mean
$m_i$. Because the state cannot move above $\bar{\theta}$, and because we impose
a maximum one-period upward movement of $J$ grid points, the realized movement
is capped at $K_i=J\wedge(n-i)$. Any probability assigned to moves larger than
$K_i$ is placed on the transition to $\theta_{i+K_i}$. The appendix gives the
exact probabilities. Each model period is one year, so a principal discount
factor of $\beta_p=0.96$.

The models are formulated in AMPL and submitted to the NEOS Server
\citep{czyzyk1998,dolan2001,gropp1997}. For the
unrestricted payment schedules, SNOPT is first run with $\mu=1000$ to obtain a
smooth starting point, after which CONOPT refines the solution with $\mu=10$.
Together, the two stages usually take less than five minutes for each case.
The restricted linear-contract problems are solved directly with CONOPT.

\subsection{Base-Case Parameter values}
The base-case calibration is summarized in Table~\ref{tab:parameters}. The state
grid is $\theta_i=i\bar{\theta}/n$. At state $\theta_i$, effort $m_i$ is the
expected number of upward grid steps before depreciation. 

\begin{table}[htbp]
\centering
\small
\begin{tabular}{ccl}
\toprule
Parameter & Value & Description\\
\midrule
$n$ & 200 & Number of grid intervals; the model has 201 states.\\
$\bar{\theta}$ & 1 & Upper endpoint of the state space; grid width is 0.005.\\
$J$ & 25 & Largest one-period upward movement in grid points.\\
$k$ & 380 & Scale of the principal's utility, $u_p(\theta)=k\theta$.\\
$h$ & 0.1 & Effort-cost coefficient in $c(m)=hm^2$.\\
$\beta_p$ & 0.96 & Principal's annual discount factor.\\
$\beta_a$ & 0.96 & Agent's discount factor, varied across experiments.\\
$q$ & 0.5 & Probability of depreciation.\\
$\delta$ & 10 & Depreciation in grid points, equal to 0.05 in state units.\\
$\mu$ & 1000, then 10 & Payment-smoothing weight used in the two-stage solve.\\
\bottomrule
\end{tabular}
\caption{Common numerical parameter settings.}
\label{tab:parameters}
\end{table}

\subsection{Equal discount factors without depreciation}
This case is the no-depreciation case: environmental quality does not
depreciate between periods.
Below are the optimal payment scheme when the principal has the same discount factor as the agent. Depending on the discount factor, the optimal payment scheme only starts paying the agent after seeing significant improvement in $\theta$.

\begin{figure}[H]
\centering
\begin{subfigure}{0.48\textwidth}
\includegraphics[width=\linewidth]{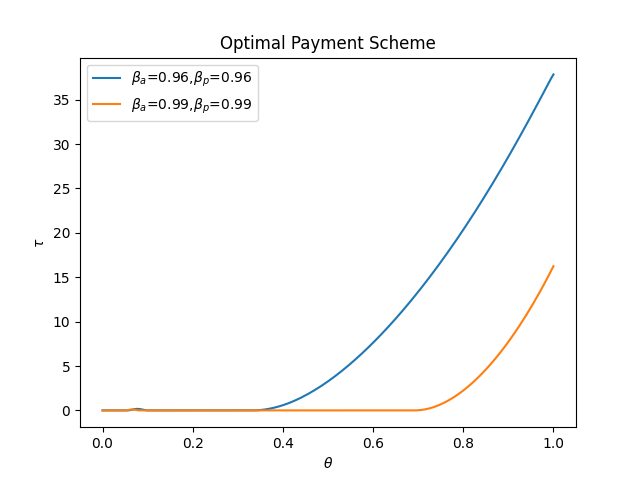}
\caption{Optimal payment.}
\end{subfigure}
\hfill
\begin{subfigure}{0.48\textwidth}
\includegraphics[width=\linewidth]{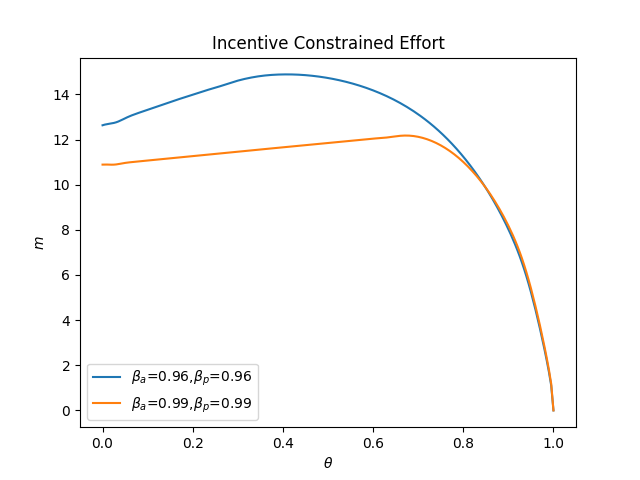}
\caption{Induced effort.}
\end{subfigure}
\caption{Optimal contract when $\beta_a=\beta_p$.}
\label{fig:same-beta}
\end{figure}

Figure~\ref{fig:same-beta} shows that increasing both discount factors from
0.96 to 0.99 delays the first substantial payment and lowers the terminal
payment. Induced effort is also lower over most states when we increase the common discount factor.,

We define the \textit{first-best} effort as the scenario in which the
incentive-compatibility constraints are removed: The principal can directly ask
the agent to exert an effort schedule $\{m_i^{FB}\}$, and the agent is fully
compensated for the cost of that effort. Thus transfers only reimburse the
agent rather than providing indirect incentives, and the effort schedule is
chosen by maximizing the principal's environmental payoff net of the fully
compensated effort cost.

\begin{figure}[H]
\centering
\includegraphics[width=0.55\linewidth]{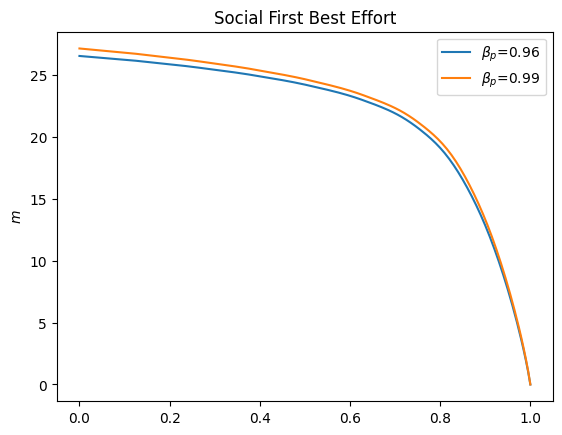}
\caption{First-best effort benchmark when $\beta_a=\beta_p$.}
\label{fig:same-beta-first-best}
\end{figure}

Figure~\ref{fig:same-beta-first-best} compares this benchmark with the
incentive-compatible effort in Figure~\ref{fig:same-beta}. The first-best
effort is much higher than the incentive-compatible effort and decreases
steadily with the state.

\subsection{Effect of depreciation}
We compare the base case with depreciation parameters $q=0.5$ and $\delta=10$, with
the no-depreciation case $\delta=0$. All other parameters are fixed at the
base values, including $\beta_a=\beta_p=0.96$. This comparison isolates how
depreciation changes the payment and effort schedules shown below.

\begin{figure}[H]
\centering
\begin{subfigure}{0.48\textwidth}
\includegraphics[width=\linewidth]{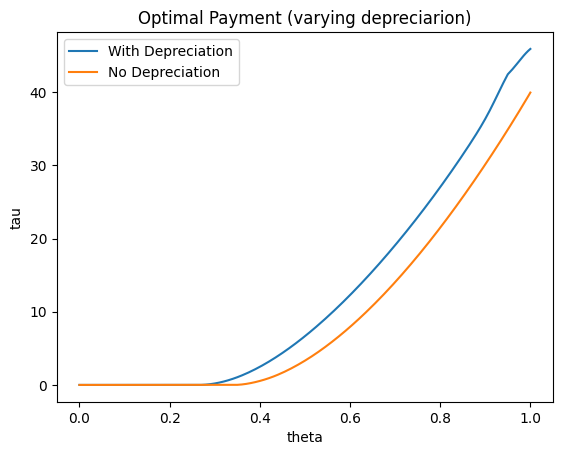}
\caption{Optimal payment.}
\end{subfigure}
\hfill
\begin{subfigure}{0.48\textwidth}
\includegraphics[width=\linewidth]{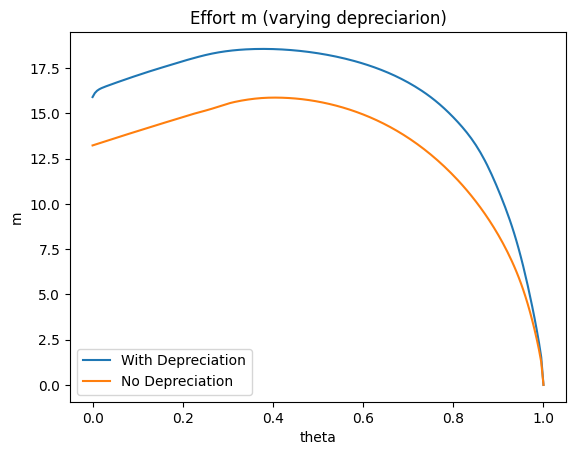}
\caption{Induced effort.}
\end{subfigure}
\caption{Optimal policy with and without depreciation.}
\label{fig:depreciation}
\end{figure}

Figure~\ref{fig:depreciation} shows that introducing depreciation leads to higher payments and higher effort over most of the state space. Because environmental improvements are less persistent, the principal has a stronger incentive to encourage continued effort.

\subsection{Effect of the agent's discount factor}
We fix $\beta_p=0.96$ and compare
$\beta_a\in\{0.8,0.9,0.96\}$, while all other parameters are the same as the base case. All cases in this
experiment include the baseline depreciation process, $q=0.5$ and $\delta=10$.
A more patient agent values future payments more strongly, so the principal can
provide incentives with a lower current payment schedule.

\begin{figure}[H]
\centering
\begin{subfigure}{0.48\textwidth}
\includegraphics[width=\linewidth]{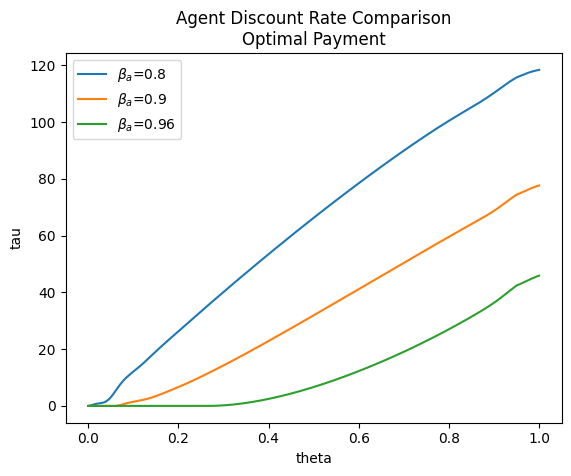}
\caption{Optimal payment.}
\end{subfigure}
\hfill
\begin{subfigure}{0.48\textwidth}
\includegraphics[width=\linewidth]{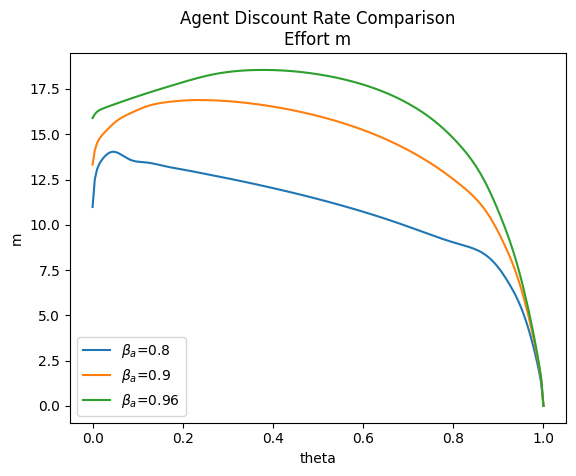}
\caption{Induced effort.}
\end{subfigure}
\caption{Effect of the agent's discount factor, holding $\beta_p=0.96$.}
\label{fig:agent-beta}
\end{figure}

Figure~\ref{fig:agent-beta} shows that the terminal payment falls when the $\beta_a$ increases, which means the agent becomes more patient. Meanwhile, the effort is also
higher over most of the state space for the more patient agent.

\subsection{Effect of the principal's utility function}
Holding $\beta_a=\beta_p=0.96$, we compare the baseline linear payoff with
\[
u_p(\theta)=k\sqrt{\theta}
\quad\text{and}\quad
u_p(\theta)=k\left(1-(1+\theta)^{-2}\right).
\]
All specifications in this experiment include the baseline depreciation
process, $q=0.5$ and $\delta=10$. We choose utility functions that are similar in scales. At $\theta=0$, all three utility functions are zero; At $\theta=1$, the inverse-square gross benefit is $0.75k$, while the other two
gross benefits equal $k$. 

\begin{figure}[H]
\centering
\begin{subfigure}{0.48\textwidth}
\includegraphics[width=\linewidth]{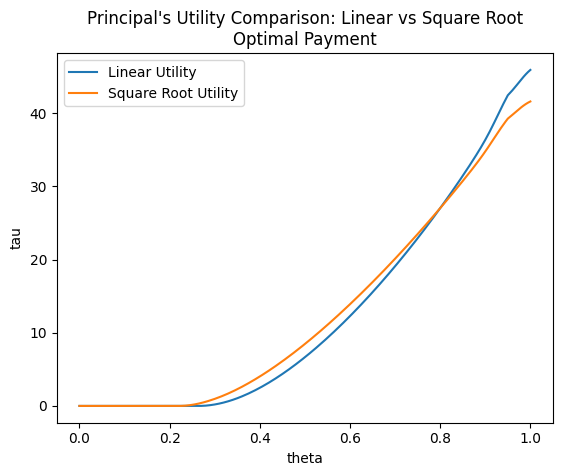}
\caption{Payment.}
\end{subfigure}
\hfill
\begin{subfigure}{0.48\textwidth}
\includegraphics[width=\linewidth]{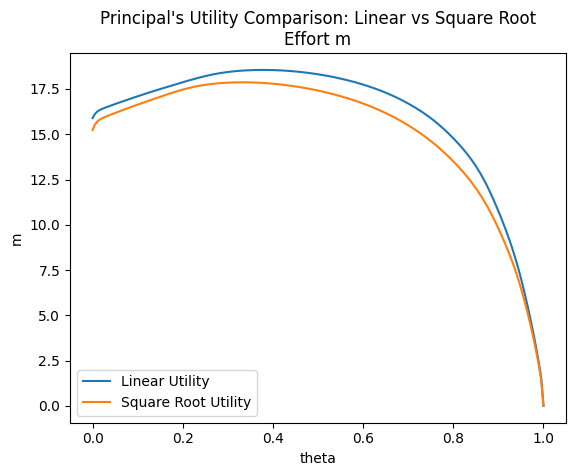}
\caption{Effort.}
\end{subfigure}
\caption{Linear versus square-root principal utility.}
\label{fig:sqrt-utility}
\end{figure}

\begin{figure}[H]
\centering
\begin{subfigure}{0.48\textwidth}
\includegraphics[width=\linewidth]{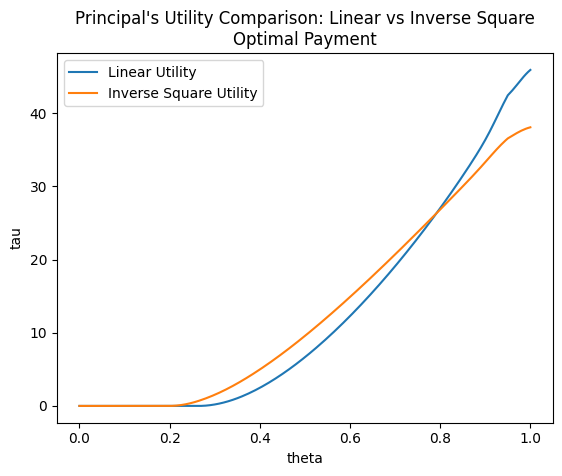}
\caption{Payment.}
\end{subfigure}
\hfill
\begin{subfigure}{0.48\textwidth}
\includegraphics[width=\linewidth]{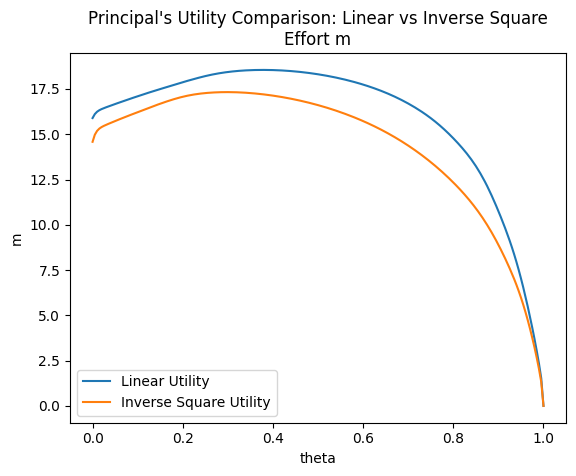}
\caption{Effort.}
\end{subfigure}
\caption{Linear versus inverse-square principal utility.}
\label{fig:inverse-utility}
\end{figure}

Both concave utility cases produce lower effort than the linear benchmark
over most of the interior. Their payments are somewhat higher at intermediate
states, cross the linear utility payment scheme around $\theta=0.8$, and are lower near the
upper boundary; the departure is larger for inverse-square utility. This
pattern is consistent with diminishing marginal environmental benefits: an
additional state improvement is less valuable at high states, weakening the
incentive to push the state toward the upper boundary.

\subsection{Flexible nonlinear versus linear contracts}
Linear contracts are simple and have approximation guarantees in the static
PES model of \citet{daili2023} and the dynamic PES model of
\citet{lomurthy2025}. Motivated by these results, we compare linear and flexible
nonlinear payment schedules to examine how contract flexibility affects
payments and induced effort in our model.
For the linear-contract experiment, the payment is restricted to
\[
\tau_i=\tau_n\frac{i}{n}.
\]
We fix $\beta_p=0.96$ and compare the payment and effort schedules under the
two contract forms for $\beta_a=0.8$ and $\beta_a=0.96$. Both contract
forms include the baseline depreciation process, $q=0.5$ and $\delta=10$.

\begin{figure}[H]
\centering
\begin{subfigure}{0.48\textwidth}
\includegraphics[width=\linewidth]{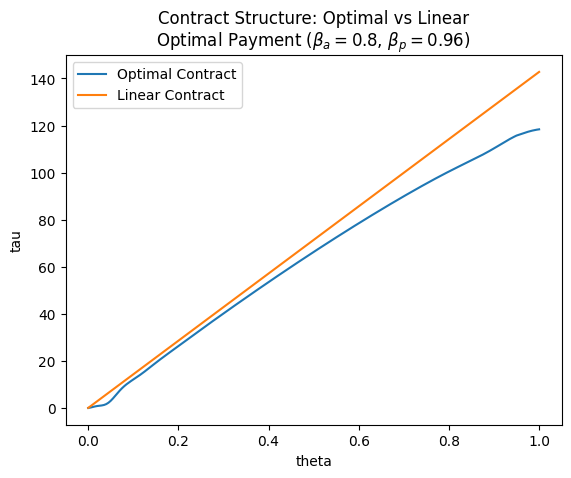}
\caption{Payment, $\beta_a=0.8$.}
\end{subfigure}
\hfill
\begin{subfigure}{0.48\textwidth}
\includegraphics[width=\linewidth]{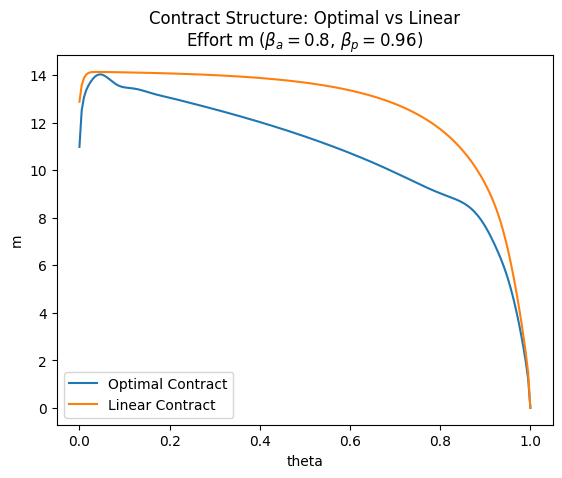}
\caption{Effort, $\beta_a=0.8$.}
\end{subfigure}

\begin{subfigure}{0.48\textwidth}
\includegraphics[width=\linewidth]{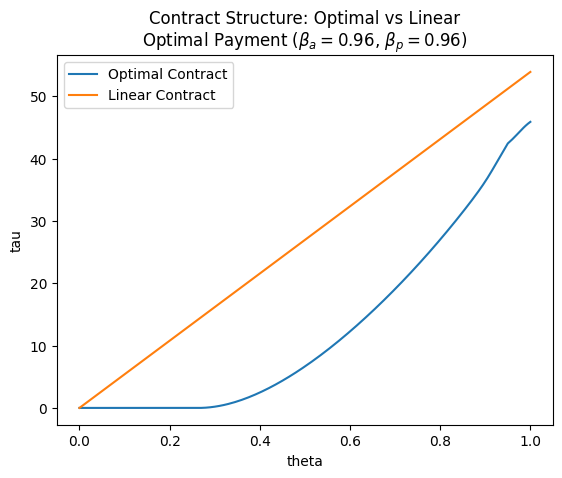}
\caption{Payment, $\beta_a=0.96$.}
\end{subfigure}
\hfill
\begin{subfigure}{0.48\textwidth}
\includegraphics[width=\linewidth]{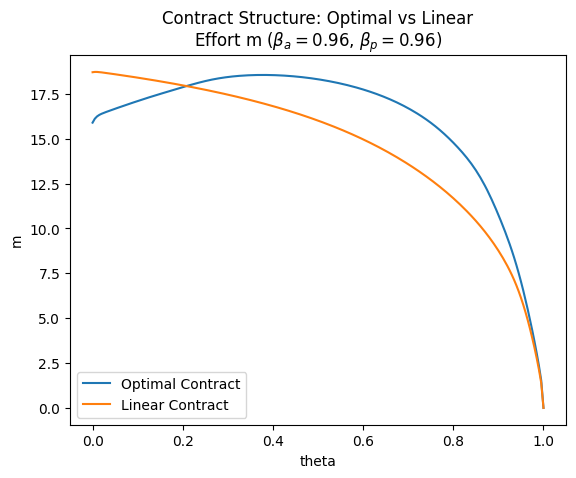}
\caption{Effort, $\beta_a=0.96$.}
\end{subfigure}
\caption{Optimal nonlinear contract versus the restricted linear contract under
two agent-discount settings.}
\label{fig:linear-contract}
\end{figure}

The linear payment schedule lies above the nonlinear schedule at nearly every
positive state in both panels, but the effort comparison depends on the
agent's patience. When $\beta_a=0.8$, the linear schedule induces higher effort
at nearly all environmental states between the lower and upper boundaries. When $\beta_a=0.96$, it induces higher effort
only below about $\theta=0.2$; the nonlinear schedule induces higher effort over
most higher states. Thus, in the latter case, the nonlinear contract delays
payments at low states and concentrates incentives at higher states.

\section{Conclusion}
This paper develops a dynamic framework for designing payments for ecosystem
services when environmental quality is observable but the effort that improves
it is hidden. The principal chooses a payment scheme over the
environmental state, while the agent responds through effort that affects the states. The formulation accommodates
different discount factors, nonlinear environmental benefits, and stochastic
depreciation.

Taken together, the experiments show that contract design depends on the
patience of both parties, the persistence of environmental gains, the shape of
environmental benefits, and the flexibility of the payment schedule. These
features determine where payments are most effective at sustaining effort.

Several extensions would strengthen the analysis. Adding individual rationality constraints and heterogeneous initial states would permit a richer analysis of enrollment and additionality. Allowing environmental quality to be observed with error would capture the imperfect monitoring common in PES programs, where payments may depend on a noisy measure of the true environmental state.
Finally, calibrating transition and depreciation parameters to a specific
restoration program would allow the model to evaluate contract duration,
permanence after payments end, and the distribution of gains between the
principal and participating landholders.

\appendix
\section{Numerical Implementation of Capped Poisson}
Let
\[
K_i := J \wedge (n-i).
\]
Instead of renormalizing the truncated Poisson distribution, we cap the transition at
$K_i$. Equivalently, if $Y \sim \mathrm{Poisson}(m)$, then the realized transition is
$\min\{Y,K_i\}$. Therefore,
\[
p_{ij}(m)
=
\begin{cases}
e^{-m}\dfrac{m^j}{j!}, & j=0,1,\ldots,K_i-1, \\[1.2em]
1-\displaystyle\sum_{\ell=0}^{K_i-1} e^{-m}\dfrac{m^\ell}{\ell!}, & j=K_i, \\[1.2em]
0, & \text{otherwise}.
\end{cases}
\]
The derivative with respect to effort $m$ is
\[
p'_{ij}(m)
=
\begin{cases}
e^{-m}\left(\dfrac{m^{j-1}}{(j-1)!}-\dfrac{m^j}{j!}\right),
& j=1,\ldots,K_i-1, \\[1.2em]
-e^{-m}, & j=0,\ K_i\geq 1, \\[1.2em]
e^{-m}\dfrac{m^{K_i-1}}{(K_i-1)!}, & j=K_i,\ K_i\geq 1, \\[1.2em]
0, & \text{otherwise}.
\end{cases}
\]
For the boundary case $K_i=0$, we have
\[
p_{i0}(m)=1,
\qquad
p'_{i0}(m)=0.
\]

\end{document}